\documentclass[journal,comsoc,10pt,twocolumn]{IEEEtran}

\usepackage{etex}

\usepackage[T1]{fontenc}

\usepackage{cite}
\ifCLASSINFOpdf
  \usepackage[pdftex]{graphicx}
  \graphicspath{{./figures/}{../pdf/}{../jpeg/}}
  \DeclareGraphicsExtensions{.pdf,.jpeg,.png,.eps,.jpg}
\else
  \usepackage[dvips]{graphicx}
  \graphicspath{{./figures/}{../eps/}}
  \DeclareGraphicsExtensions{.pdf,.jpeg,.png,.eps,.jpg}
\fi
\usepackage{epstopdf}

\usepackage{caption}

\usepackage{tikz}
\usetikzlibrary{automata,positioning,chains,shapes,arrows}
\usepackage{pgfplots}
\usetikzlibrary{plotmarks}
\newlength\fheight
\newlength\fwidth
\pgfplotsset{compat=newest}
\pgfplotsset{plot coordinates/math parser=false}

\usepackage{amsmath}
\usepackage{bm}
\usepackage{dsfont}
\usepackage{amssymb}
\usepackage{pifont}
\newcommand{\cmark}{\ding{51}}%
\newcommand{\xmark}{\ding{55}}%
\newcommand{\LO}{LoRa\textsuperscript{\texttrademark}\,}
\newcommand{\IN}{Ingenu\textsuperscript{\texttrademark}\,}
\newcommand{\SF}{SIGFOX\textsuperscript{\texttrademark}\,}
\newcommand{\Fig}[1]{Fig.~\ref{#1}}
\usepackage{algorithmic}
\usepackage{array}
\newcolumntype{L}[1]{>{\raggedright\let\newline\\\arraybackslash\hspace{0pt}}m{#1}}
\newcolumntype{C}[1]{>{\centering\let\newline\\\arraybackslash\hspace{0pt}}m{#1}}
\newcolumntype{R}[1]{>{\raggedleft\let\newline\\\arraybackslash\hspace{0pt}}m{#1}}

\usepackage{tabularx}
\usepackage{caption}
\usepackage{multirow}
\usepackage{multicol}
\usepackage{booktabs}
\usepackage[applemac]{inputenc}

\ifCLASSOPTIONcompsoc
  \usepackage[caption=false,font=normalsize,labelfont=sf,textfont=sf]{subfig}
\else
  \usepackage[caption=false,font=footnotesize]{subfig}
\fi
\usepackage{dblfloatfix}
\ifCLASSOPTIONcaptionsoff
  \usepackage[nomarkers]{endfloat}
 \let\MYoriglatexcaption\caption
 \renewcommand{\caption}[2][\relax]{\MYoriglatexcaption[#2]{#2}}
\fi
\let\MYorigsubfloat\subfloat
\renewcommand{\subfloat}[2][\relax]{\MYorigsubfloat[]{#2}}
\usepackage{url}
\usepackage{color}

\usepackage{verbatim}

\usepackage{microtype}
\DisableLigatures[f]{encoding=*,family=*}

\usepackage{siunitx}
\usepackage{soul}

\begin{document}
% According to [The Chicago Manual of Style, 16th Ed., Par. 8.155],
% the general rules are the following ones:
% 1) Capitalize the first and the last word;
% 2) Capitalize nouns, pronouns, adjectives, verbs, adverbs, and subordinate conjunctions;
% 3) Lowercase articles (a, an, the), coordinating conjunctions, and prepositions;
% 4) Lowercase the "to" in an infinitive (I want to play guitar).
%\title{A New Paradigm for Internet of Things and Smart Cities Connectivity: Long Range Communications in Unlicensed Bands}
\title{Long-Range Communications in Unlicensed Bands: the Rising Stars in the IoT and Smart City Scenarios}
%\title{Long Range Wide Area Networks: the Final Solution to IoT and Smart City Connectivity?}


\author{
        Marco~Centenaro,~\IEEEmembership{Student~Member,~IEEE,}
	Lorenzo~Vangelista,~\IEEEmembership{Senior~Member,~IEEE,}
        Andrea~Zanella,~\IEEEmembership{Senior~Member,~IEEE,}
        and~Michele~Zorzi,~\IEEEmembership{Fellow,~IEEE}% <-this % stops a space
\thanks{The authors are with the Department of Information Engineering, University of Padova, Italy; e-mail: \{{\tt firstname.lastname\}@dei.unipd.it}. % <-this % stops a space
%{\tt lorenzo.vangelista@unipd.it}, {\tt \{marco.centenaro,zanella,michele.zorzi\}@dei.unipd.it}.}% <-this % stops a space
 L. Vangelista and M. Zorzi are also with Patavina Technologies s.r.l., Padova, Italy; e-mail: {\tt \{firstname.lastname\}@patavinatech.com}, web: {\tt http://www.patavinatech.com/en/}.}% <-this % stops a space
%\thanks{Manuscript received Month~Day, Year; revised Month~Day, Year.}
\thanks{This paper is partly based on the paper ``Long-range IoT technologies: the dawn of
LoRa\textsuperscript{\texttrademark}\,'' by L. Vangelista, A. Zanella and M. Zorzi presented at {\em Fabulous 2015} conference, Sept. 23--25, 2015, Ohrid, Republic of Macedonia.}}


\begin{verbatim}
This article has been accepted for publication in the October 2016 issue of 
IEEE Wireless Communications, Special Issue on 
``ENABLING WIRELESS COMMUNICATION AND NETWORKING TECHNOLOGIES FOR INTERNET OF THINGS''

PLEASE, CITE THE PAPER AS FOLLOWS:

Plain text:
M. Centenaro, L. Vangelista, A. Zanella, and M. Zorzi, ?Long-Range Communications in
Unlicensed Bands: the Rising Stars in the IoT and Smart City Scenarios,? IEEE
Wireless Communications, Vol. 23, Oct. 2016.

BibTex:
@ARTICLE{CVZZ16,
author={M. Centenaro and L. Vangelista and A. Zanella and M. Zorzi},
journal={IEEE Wireless Communications},
title={{Long-Range Communications in Unlicensed Bands: the Rising Stars in the IoT
and Smart City Scenarios}},
year={2016},
volume={23},
keywords={Internet of Things;Smart Cities;Low-Power Wide Area Network (LPWAN);LoRa;
SIGFOX;Ingenu;Cellular IoT},
month={October},}

\end{verbatim}
\clearpage
\maketitle
\begin{abstract}
Connectivity is probably the most basic building block of the Internet of Things (IoT) paradigm.
Up to know, the two main approaches to provide data access to the \emph{things} have been based either on multi-hop mesh networks using short-range communication technologies in the unlicensed spectrum, or on long-range, legacy cellular technologies, mainly 2G/GSM, operating in the corresponding licensed frequency bands.
Recently, these reference models have been challenged by a new type of wireless connectivity, characterized by low-rate, long-range transmission technologies in the unlicensed sub-GHz frequency bands, used to realize access networks with star topology which are referred to a \emph{Low-Power Wide Area Networks} (LPWANs).
In this paper, we introduce this new approach to provide connectivity in the IoT scenario, discussing its advantages over the established paradigms in terms of efficiency, effectiveness, and architectural design, in particular for the typical Smart Cities applications.
\end{abstract}

\begin{IEEEkeywords}
Internet of Things, Smart Cities, Low-Power Wide Area Network (LPWAN), \LO, SIGFOX\textsuperscript{\texttrademark}, Ingenu\textsuperscript{\texttrademark}, Cellular~IoT.
\end{IEEEkeywords}

\begin{picture}(2,10)(0,-370)
\put(-10,0){\footnotesize{A revised version of this manuscript will appear in IEEE Wireless Communications, Oct. 2016}}
\end{picture}

% \fi
\IEEEpeerreviewmaketitle

%%%%%%%%%%%%%%%%%%%%%%%%%%%%%%%%%%%%%%%%%%%%%%%%%
%%%%%%%%%%%%%%%%%%%%%%%%%%%%%
\section{Introduction}\label{intro}
%%%%%%%%%%%%%%%%%%%%%%%%%%%%%
The Internet of Things (IoT) paradigm refers to a {\em network} of interconnected {\em things}.
The {\em network} is normally intended as the IP network and the {\em things} are devices, such as sensors and/or actuators, equipped with a telecommunication interface and with processing and storage units.
This communication paradigm should hence enable seamless integration of potentially any object into the Internet, thus allowing for new forms of interactions between human beings and devices, or directly between device and device, according to what is commonly referred to as the Machine-to-Machine (M2M) communication paradigm \cite{M2M1}. 

The development of the IoT is an extremely challenging topic and the debate on how to put it into practise is still open. The discussion interests all layers of the protocol stack, from the physical transmission up to data representation and service composition.
However, the whole IoT castle rests on the wireless technologies that are used to provide data access to the end devices. 

For many years, multi-hop short-range transmission technologies, such as ZigBee and Bluetooth, have been considered a viable way to implement IoT services \cite{IoT1,IoT2,IoT3}.
Although these standards provide very low power consumption, which is a fundamental requirement for many IoT devices like, e.g., smart sensors, their limited coverage constitutes a major obstacle, in particular when the application scenario involves services that require urban-wide coverage, as in typical Smart City applications \cite{IoT3}. 
The experimentations of some initial Smart Cities services have, indeed, revealed the limits of the multi-hop short-range paradigm for this type of IoT applications, stressing the need for an access technology that can allow for a \emph{place-\&-play} type of connectivity, i.e., that makes it possible to connect any device to the IoT by simply placing it in the desired location and switching it on \cite{M2M4}.

In this perspective, wireless cellular networks may play a fundamental role in the spread of IoT, since they are able to provide  ubiquitous and transparent coverage \cite{M2M1,M2M2,M2M3}.
In particular, the Third Generation Partnership Project (3GPP), which is the standardization body for the most important cellular technologies, is attempting to revamp 2G/GSM to support IoT traffic, implementing the so-called Cellular~IoT (CIoT) architecture \cite{CIoT}.
On the other side, the latest cellular network standards, e.g., UMTS and LTE, were not designed to provide machine-type services to a massive number of devices.
In fact, differently from traditional broadband services, IoT communication is expected to generate, in most cases, sporadic transmissions of short packets.
At the same time, the potentially huge number of IoT devices asking for connectivity through a single Base Station (BS) would raise new issues related to the signaling and control traffic, which may become the bottleneck of the system \cite{M2M4}. %,zanella_m2m}.
All these aspects make current cellular network technologies not suitable to support the envisioned IoT scenarios, while, on the other hand, a number of research challenges still need to be addressed before the upcoming 5G cellular networks may natively support IoT services.

A promising alternative solution, standing in between short-range multi-hop technologies operating in the unlicensed industrial, scientific, and medical (ISM) frequency bands, and long-range cellular-based solutions using licensed broadband cellular standards, is provided by the so-called \emph{Low-Power Wide Area Networks} (LPWANs).

These kinds of networks exploit sub-GHz, unlicensed frequency bands and are characterized by long-range radio links and star topologies. The end devices, indeed, are directly connected to a unique collector node, generally referred to as \emph{gateway}, which also provides the bridging to the IP world.
The architecture of these networks is designed to provide wide area coverage and ensure the connectivity also to nodes that are deployed in very harsh environments.

The goal of this paper is to provide an introductory overview of the LPWAN paradigm and of its main technological interpretations.
We will discuss the advantages provided by this new type of connectivity with respect to the more traditional solutions operating in the unlicensed spectrum, especially for applications related to Smart Cities.
To substantiate our argumentation, we will refer to some preliminary experiments and deployments of IoT networks based on \LO, one of the LPWAN solutions available on the market today.

The rest of the paper is organized as follows.
In Section~\ref{state}, current wireless technologies and service platforms for the IoT connectivity are reviewed.
The potential of LPWANs is discussed is Section~\ref{paradigm}, while Section~\ref{review} describes the commercial LPWAN products available today, focusing in greater detail on \LO, whose characteristics make it a good representative of the LPWAN family, while its open specifications make it possible to access some details of its most interesting and specific mechanisms.
In Section \ref{experimental} we discuss the experience gained with some experimental deployments of a \LO network.
Conclusions and final remarks can be found in Section~\ref{conclusions}.

%State of the art
%%%%%%%%%%%%%%%%%%%%%%%%%%%%%
\section{A Quick Overview of Current IoT Communication Standards}\label{state}
%%%%%%%%%%%%%%%%%%%%%%%%%%%%%

%The Internet of Things paradigm refers to a {\em network} of interconnected {\em things}.
%The {\em network} is normally intended as the IP network and the {\em things} as sensors and/or actuators equipped with a telecommunication interface.
%What is normally not highlighted enough is that the type of connection is a wireless short range and - usually - not IP-based one.
Although the IoT paradigm does not set any constraint on the type of technology used to connect the end devices to the Internet, it is a fact that wireless communication is the only feasible solution for a large majority of the IoT applications and services.
As mentioned, the current practice considers either cellular-based or multi-hop short-range technologies.
In the latter case, the connected {\em things} usually run on dedicated protocol stacks, suitably designed to cope with the constraints of the end devices.
Furthermore, at least one such device is required to be connected to the IP network, acting as \emph{gateway} for the other nodes.
The architecture is hence distributed, with many ``islands'' (sub-nets) that operate according to different connectivity protocols, and are connected to the IP network via gateways.
The applications and services are deployed on top of this connectivity level, according to a {\em distributed service} layer.
The applications may run either locally, i.e., in the sub-net, or, more and more often (as typical in the Smart City scenario), using cloud computing services.

At this level we can find the {\em IoT platforms} that act as a unifying framework, enabling the service creation and delivery, as well as the operation, administration, and maintenance of the {\em things} and the gateways.
%The market for the IoT platforms is booming right now, for sure because of the need of service creation, but also because they fulfil the need for interoperability between the different sub-nets.
Nowadays, the most important ``de facto'' standards in the IoT arena are the following:
\begin{enumerate}
\item extremely short-range systems, e.g., Near Field Communications (NFC) enabled devices;
\item short-range passive and active Radio Frequency IDentification (RFID) systems;
\item systems based on the family of IEEE 802.15.4 standards like ZigBee\textsuperscript{\texttrademark}, 6LoWPAN, Thread-based systems;
\item Bluetooth-based systems, including Bluetooth Low Energy (BLE); 
\item proprietary systems, including Z-Wave\textsuperscript{\texttrademark}, CSRMesh\textsuperscript{\texttrademark}, i.e., the Bluetooth mesh by Cambridge Silicon Radio (a company now owned by Qualcomm), EnOcean\textsuperscript{\texttrademark}; 
\item systems mainly based on IEEE 802.11/Wi-Fi\textsuperscript{\texttrademark}, e.g., those defined by the ``AllSeen Alliance'' specifications, which explicitly include the gateways, or by the ``Open Interconnect Consortium''. 
\end{enumerate}

The vast majority of the connected {\em things} at the moment is using IEEE 802.15.4-based systems, in particular ZigBee\textsuperscript{\texttrademark}.
%\MCE{Trovare una reference per avvallare questa asserzione!}
The most prominent features of these networks are that they operate mainly in the $2.4$~GHz and optionally in the $868/915$~MHz unlicensed frequency bands and the network level connecting these {\em nodes}\footnote{{\em Node} is a term that is frequently used to indicate a connected {\em thing}, with emphasis on the communication part.} uses a mesh topology.
The distances between the nodes in this kind of systems ranges from few meters, up to roughly $100$~meters, depending on the surrounding environment (presence of walls, obstacles, and so on).

To better appreciate the comparison with LPWAN technologies, it is worth highlighting the main characteristics of these IoT technologies.
\begin{itemize}
\item \emph{Mesh networking}. Multihop communication is necessary to extend the network coverage beyond the limited reach of the low-power transmission technology used. Furthermore, the mesh architecture can provide resilience to the failure of some nodes. On the other hand, the maintenance of the mesh network requires non-negligible control traffic, and multi-hop routing generally yields long communication delays, and unequal and unpredictable energy consumption among the devices. 
\item \emph{Short coverage range - high data rate}. The link level technologies used in these systems tend to privilege the data rate rather than the sensitivity, i.e., in order to recover from the network delays due to the mesh networking, these networks have a relatively high raw link bit rate (e.g., 250 Kbit/s), but they are not robust enough to penetrate building walls and other obstacles (even in the 868/915 MHz band). In other words, in the trade-off between rate and sensitivity, the rate is usually preferred.
\end{itemize}

%%%%%%%%%%%%%%%%%%%%%%%%%%%%%%%%%%%%%%%%%%%%%%%%%%%%
% The new paradigm: long range IoT communications in unlicensed bands
%%%%%%%%%%%%%%%%%%%%%%%%%%%%%%%%%%%
\section{A New Paradigm: Long-Range IoT Communications in Unlicensed Bands}\label{paradigm}
%%%%%%%%%%%%%%%%%%%%%%%%%%%%%%%%%%%

%In the previous section, we have seen that the current IoT systems is mainly characterized by a mesh topology and short range connections with moderately high rate.
As a counterpart of the unlicensed short-range technologies for the IoT mentioned in the previous sections, we turn our attention to the emerging paradigm of LPWAN. 

Most LPWANs operate in the unlicensed ISM bands centered at 2.4 GHz, 868/915 MHz, 433 MHz, and 169 MHz, depending on the region of operation. 
The radio emitters operating in these frequency bands are commonly referred to as ``Short Range Devices'' \cite{etsi300}, a rather generic term that delivers the idea of coverage ranges of few meters, which was indeed the case for the previous ISM wireless systems.
Nonetheless, the ERC Recommendation 70-03 specifies that \textsf{``The term \emph{Short Range Device} (SRD) is intended to cover the radio transmitters which provide either uni-directional or bi-directional communication which have low capability of causing interference to other radio equipment.''}
Therefore, there is no explicit mention of the actual coverage range of such technologies. 

LPWAN solutions are indeed examples of ``short-range devices'' with cellular-like coverage ranges, in the order of 10--15~km in rural areas, and 2--5~km in urban areas. This is possible thanks to a radically new physical layer design, aimed at very high receiver sensitivity.
For example, while the nominal sensitivity of ZigBee\textsuperscript{\texttrademark} and Bluetooth receivers is about -125 dBm and -90~dBm, respectively, the typical sensitivity of a LPWAN receiver is around -150~dBm (see Section~\ref{review}).

%For example, the modulation of a prominent LPWAN technology called \LO (see Section~\ref{review} for further details), operating in the 868 MHz band, is designed in such a way that its receiver has a sensitivity of -148~dBm.
%We recall that the sensitivity of a ZigBee\textsuperscript{\texttrademark} receiver is about -125 dBm, while that one of a Bluetooth receiver is about -90~dBm. 

The downside of these long-range connections is the low data rate, which usually ranges from few hundred to few thousand bit/s, significantly lower than the bitrates supported by the actual short-range technologies, e.g., 250~Kbit/s in ZigBee\textsuperscript{\texttrademark} and 1--2~Mbit/s in Bluetooth. 
%In fact, in accordance with the above quotations, all radio emitters operating in the ISM bands are subject to specific limitations in the use of the spectrum, in order to limit the interference produced to other devices. 
%For example, the \LO system has a raw physical layer data rate ranging from few hundred of bits/s to few Kbit/s compared to 250~Kbit/s of ZigBee\textsuperscript{\texttrademark} systems. 
%The question that is immediately raised is whether the LPWAN, with such a low bit rate, is suitable for the IoT applications.
%The answer, which will be further discussed in Section \ref{conclusions}, is positive. 
%However a first remark should be done to set the right framework for comparison.
However, because of the signaling overhead and the multi-hop packet forwarding method, the actual flow-level throughput provided by such  short-range technologies is generally much lower than the nominal link-layer bitrate, settling to values that are comparable to those reached by the single-hop LPWANs. 
While such low bitrates are clearly unsatisfactory for most common data-hungry network applications, many Smart City and IoT services are expected to generate a completely different pattern of traffic, characterized by sporadic and intermittent transmissions of very small packets (typical of monitoring and metering applications, remote switching control of equipment, and so on).
Furthermore, many of these applications are rather tolerant to delays and packet losses and, hence, are suitable for the connectivity service provided by LPWANs. 

%The main reason why the LPWAN low rate is not an issue is that the vast majority of such applications has a sporadic and intermittent traffic with very small payloads\footnote{One should think of applications in the area of monitoring and metering, switching on and off some equipment, etc.}

%If we want to compare the data rate of LPWAN with that of, e.g., ZigBee or 6LoWPAN, we should actually take as a reference the actual data rate between two nodes in a ZigBee\textsuperscript{\texttrademark} or 6LoWPAN {\em network} and compare it with that of a single link in the same networks: we could, then, realize that, for ZigBee\textsuperscript{\texttrademark} or 6LoWPAN systems, this is very often on the order of 10-20 Kbit/s, i.e., the overhead of the mesh architecture reduces by one order of magnitude the actual rate with respect to the single link bit rate.
%The main reason why the LPWAN low rate is not an issue is that the vast majority of such applications has a sporadic and intermittent traffic with very small payloads\footnote{One should think of applications in the area of monitoring and metering, switching on and off some equipment, etc.}
% A set of target applications of utmost importance is the Smart City scenario, which are not affected by this limitation on the data rate. 

Another important characteristic of LPWANs is that the {\em things}, i.e., the end devices, are connected directly to one (or more) gateway with a single-hop link, very similar to the classic cellular network topology.
This greatly simplifies the coverage of large areas, even nation-wide, by re-using the existing infrastructure of the cellular networks.
For example, \LO systems are being deployed by telecommunication operators like Orange and Bouygues Telecom in France, by Swisscom in Switzerland, and by KPN in the Netherlands, while \SF has already deployed a nation-wide access network for M2M and IoT devices in many central European countries, from Portugal to France. %In all of these deployment the target is to have a nationwide coverage.
Furthermore, the star topology of LPWANs makes it possible to have greater control of the connection latency, thus potentially enabling the support of interactive applications that require predictable response times such as, for example, the remote control of street lights in a large city, the operation of barriers to limited-access streets, the intelligent control of traffic lights, and so on. 

Besides the access network, the similarity between LPWANs and legacy cellular systems further extends to the bridging of the technology-specific wireless access to the IP-based packet switching core network.
Indeed, the LPWAN gateways play a similar role as the Gateway GPRS Support Node (GGSN) in GPRS/UMTS networks, or the Evolved Packet Core in LTE, acting as point-of-access for the end devices to the IP-based core network and forwarding the data generated by \emph{things} to a logic controller, usually named \emph{Network Server}. 

Therefore, LPWANs inherit the basic aspects of the legacy cellular systems architecture that, however, is stripped of most advanced features, such as the management of user mobility and resource scheduling.
The combination of the simple but effective topology of cellular systems with a much lighter management plane, makes the LPWAN approach particularly suitable to support services with relatively low Average Revenue Per User, such as those envisioned in the Smart City scenario.

%One should note that, in the context of LPWAN, the BSs are often called {\em gateways}: this reflects the fact that the connectivity from/to the {\em things} and the BS is a particular air interface which is then connected to the IP-based core network connecting the gateways, once again in a star topology, to a {\em network server}.
%This network server -- in the same way as the GGSN (Gateway GPRS Support Node) in 2G networks -- is the connecting entity to the public Internet.

%It is clear from the above discussion that the LPWAN is much more suitable for Smart City applications like street lighting, waste management, street parking management, etc. 

%%%%%%%%%%%%%%%%%%%%%%%%%%%%%%%%%%%%%%%%%%%%%%%%%%%%
% Review of long range IoT communications systems in unlicensed bands
%%%%%%%%%%%%%%%%%%%%%%%%%%%%%%%%%%
\section{A Review of Long-Range IoT Communications Systems in Unlicensed Bands}\label{review}
%%%%%%%%%%%%%%%%%%%%%%%%%%%%%%%%%%

%\LVA{fare una carrellata e una comparazione fra modulazioni a banda stretta e banda larga, SigFox, Lora e OnRamp}

In this section we quickly overview three of the most prominent technologies for LPWANs, namely \SF, \IN, and \LO. In particular, we will describe in greater detail the \LO technology, which is gaining more and more momentum, and whose specifications are publicly available, thus making it possible to appreciate some of the technical choices that characterize LPWAN solutions.
In Tab.~\ref{tab:standards_comparison} a comparison between these LPWAN radio technologies can be found.

\subsection{\bf \SF}
\SF,\footnote{\url{http://www.sigfox.com}} the first LPWAN technology proposed in the IoT market, was founded in 2009 and has been growing very fast since then. %\cite{sigfox}
The \SF physical layer employs aa Ultra Narrow Band (UNB) wireless modulation, while the network layer protocols are the ``secret sauce'' of the \SF network and, as such, there exists basically no publicly available documentation. Indeed, the \SF  business model is that of an operator for IoT services, which hence does not need to open the specifications of its inner modules. 

The first releases of the technology only supported uni-directional uplink communication, i.e., from the device towards the aggregator; however bi-directional communication is now supported.
\SF claims that each gateway can handle up to a million connected objects, with a coverage area of 30--50~km in rural areas and 3--10~km in urban areas. 

\subsection{\bf \IN}
An emerging star in the landscape of LPWANs is \IN, a trademark of On-Ramp Wireless, a company headquartered in San~Diego (USA).\footnote{\url{http://www.onrampwireless.com}} % \cite{onramp}.
On-Ramp Wireless has been pioneering the 802.15.4k standard \cite{8021514k}.
The company developed and owns the rights of the patented technology called Random Phase Multiple Access (RPMA\textsuperscript{{\textregistered}}) \cite{onramp-patent}, which is deployed in different networks.
Conversely to the other LPWAN solutions, this technology works in the 2.4~GHz band but, thanks to a robust physical layer design, can still operate over long-range wireless links and under the most challenging RF environments.

\subsection{\bf The \LO System}\label{lora}
\LO is a new physical layer LPWAN solution, which has been designed and patented by Semetch Corporation that also manufactures the chipsets \cite{Sforza2013}. 
More specifically, the PHY is a derivative of \emph{Chirp Spread Spectrum} (CSS)  \cite{Berni1973}, where the innovation consists in ensuring the phase continuity between different chirp symbols in the preamble part of the physical layer packet, thus enabling a simpler and more accurate timing and frequency synchronization, without requiring expensive components that generate a stable local clock in the \LO node. 

The technology employs a spreading technique, according to which a symbol is encoded in a \emph{longer sequence of bits}, thus reducing the signal to noise and interference ratio required at the receiver for correct reception, without changing the frequency bandwidth of the wireless signal.
The length of the spreading code can be varied, thus making it possible to provide variable data rates, giving the possibility to trade throughput for coverage range, or link robustness, or energy consumption. 

The system has been designed to work in the 169~MHz, 433~MHz and 915~MHz bands in the USA, but in Europe it works in the 868~MHz band.
According to the regulation in \cite{etsi300220}, the radio emitters are required to adopt duty cycled transmission (1\% or 0.1\%, depending on the sub-band), or the so-called Listen Before Talk (LBT) Adaptive Frequency Agility (AFA) technique, a sort of carrier sense mechanism used to prevent severe interference among devices operating in the same band.
According to the specification in \cite{lorawan}, \LO (as well as \SF) uses the duty cycled transmission option only, which limits the rate at which the end device can actually generate messages.
However, by supporting multiple channels, \LO makes it possible for an end node to engage in longer data exchange procedures by changing carrier frequency, while respecting the duty cycle limit in each channel. 

While the PHY layer of \LO is proprietary, the rest of the protocol stack, known as LoRaWAN\textsuperscript{{\texttrademark}}, is kept open, and its development is carried out by the \LO Alliance,\footnote{\url{https://www.lora-alliance.org/}} led by IBM, Actility, Semtech, and Microchip. %\cite{lora-alliance}

\begin{figure}[t!]
\begin{center}
\includegraphics[width=0.9\columnwidth]{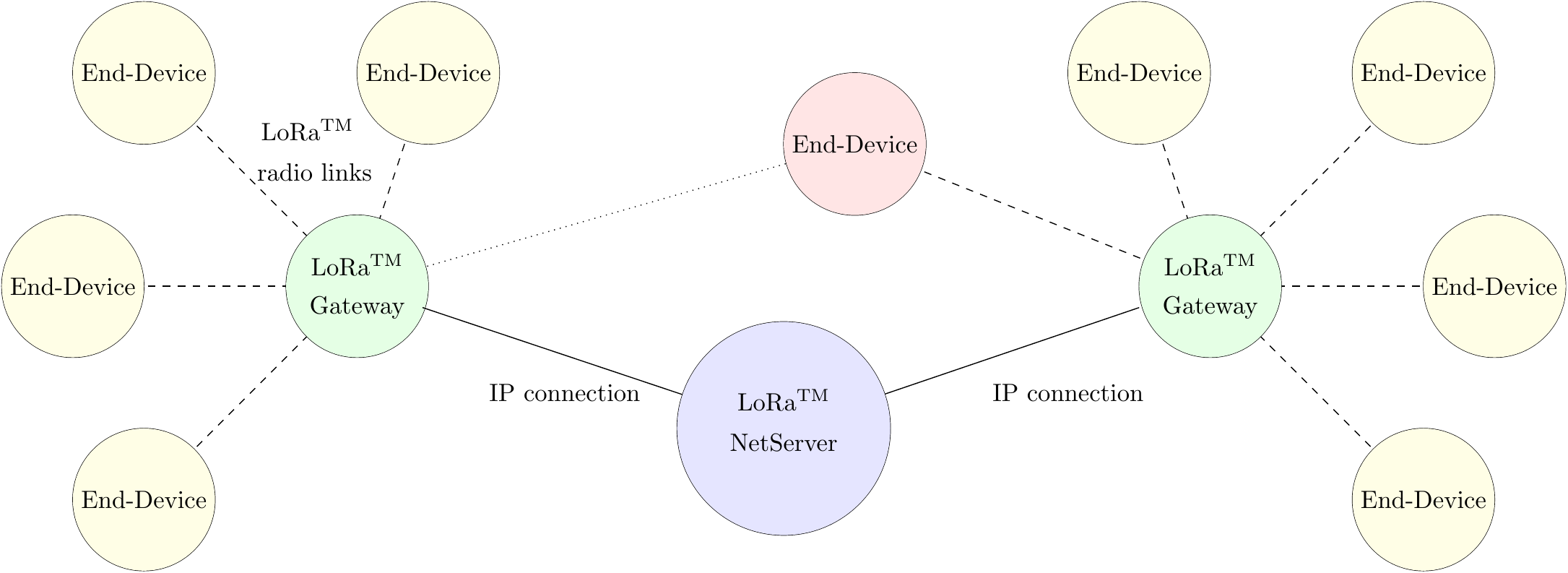}
\caption{\LO system architecture}
\label{architecture}
\end{center}
\end{figure}

%The \LO system is composed of three main components: 
%\begin{itemize}
%\item \LO End-devices: sensors/actuators connected via the \LO radio interface to one or more \LO Gateways;
%\item \LO Gateways: concentrators that bridge end devices to the central element of of the network architecture, called \LO NetServer;
%\item\LO NetServer: the server that manages the whole network (radio resource management, admission control, security, etc). 
%\end{itemize}
As exemplified in Fig.~\ref{architecture}, the \LO network is typically laid out in a {\em star-of-stars} topology, where the end devices are connected via a single-hop \LO link to one or many gateways that, in turn, are connected to a common Network Server (NetServer) via standard IP protocols.

\begin{figure}[t!]
\begin{center}
\includegraphics[width=0.9\columnwidth]{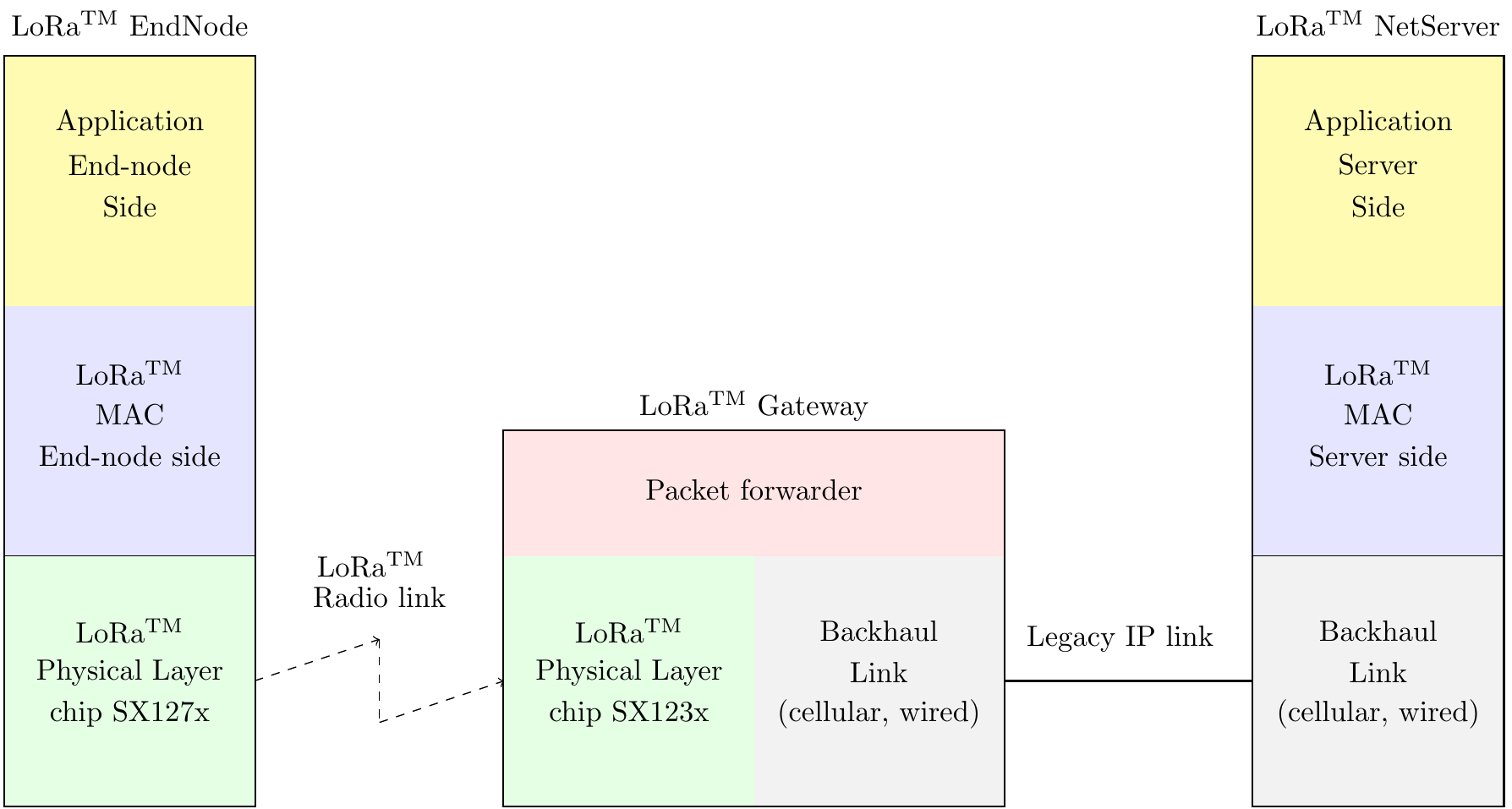}
\caption{\LO protocol architecture.}
\label{protocol}
\end{center}
\end{figure}

The gateways relay messages between the end devices and the NetServer according to the protocol architecture represented in \Fig{protocol}.
Conversely to standard cellular network systems, however, the end devices are not required to associate to a certain gateway to get access to the network, but only to the NetServer.
The gateways act as a sort of relay/bridge and simply forward to their associated NetServer all successfully decoded messages sent by any end device, after adding some information regarding the quality of the reception.
The NetServer is hence in charge of filtering duplicate and unwanted packets, and of replying to the end devices by choosing one of the in-range gateways, according to some criterion (e.g., better radio connectivity).
The gateways are thus totally transparent to the end devices, which are logically connected directly to the NetServer.
Note that current full-fledged \LO Gateways allow for the parallel processing of up to 9 \LO channels, where a channel is identified by a specific sub-band and spreading factor.

This access mode greatly simplifies the management of the network access for the end nodes, moving all the complexity to the NetServer.
Furthermore, the end nodes can freely move across cells served by different gateways without generating any additional signaling traffic in the access network, nor in the core network.
Finally, we observe that increasing the number of gateways that serve a certain end device will increase the reliability of its connection to the NetServer, which may be interesting for critical applications. 

A distinguishing feature of the \LO network is that it envisages three classes of end devices, named \emph{Class A} (for \emph{All}), \emph{Class B} (for \emph{Beacon}) and \emph{Class C} (for \emph{Continuously listening}), each associated to a different operating mode \cite{lorawan}.

Class A defines the default functional mode of the \LO networks, and must be mandatorily supported by all \LO devices.
In a Class A network, transmissions are always initiated by the end devices, in a totally asynchronous manner.
After each uplink transmission, the end device will open (at least) two reception windows, waiting for any command or data packet returned  by the NetServer.
The second window is opened on a different sub-band (previously agreed with the NetServer) in order to increase the resilience against channel fluctuations.
Class A networks are mainly intended for monitoring applications, where the data which are produced by the end devices have to be collected by a control station. 

Class B has been introduced to decouple uplink and downlink transmissions.
Class B end devices, indeed, synchronize with the NetServer by means of beacon packets which are broadcast by Class B gateways and can hence receive downlink data or command packets in specific time windows, irrespective of the uplink traffic.
Therefore, Class B is then intended for end devices that need to receive commands from a remote controller, e.g., switches or actuators.

Finally, Class C is defined for end devices without (strict) energy constraints (e.g., connected to the power grid), which can hence keep the receive window always open.

It is worth noting that, at the time of writing, Class A and B specifications are provided in \cite{lorawan}, while Class C specifications are still in draft form.


%\subsubsection{\LO Physical Layer}
%\label{PHY}

%\subsubsection{\LO MAC}
The MAC layer, according to LoRaWAN\textsuperscript{{\texttrademark}} specification \cite{lorawan}, is basically an ALOHA protocol  controlled primarily by the LoRa NetServer.
A description of the protocol is beyond the scope of this paper and can be found in \cite{lorawan}. 
%A distinguishing feature of the \LO MAC is the {\em Adaptive Data Rate} (ADR), which allows the NetServer to adapt the transmit rate of an end device by changing the spreading factor, in order to find the best trade-off between energy efficiency and link robustness.
Overall the \LO MAC has been designed attempting to mimic as much as possible the IEEE 802.15.4 MAC.
The objective is to simplify the accommodation, on top of the \LO MAC, of the major protocols now running on top of the IEEE 802.15.4 MAC, such as 6LoWPAN and CoAP. 
A clear analogy is the {\em authentication} mechanism, which is taken directly from the IEEE 802.15.4 standard using the 4-octet $MIC$ (Message Integrity Code).

\begin{table*}[t]
   	\caption[]{Comparison between LPWAN radio technologies.}
   	\label{tab:standards_comparison}
   	\centering
   	\begin{tabular}{| c | c | c | c |}
   	\hline
								& \textbf{\SF}		& \textbf{\IN}	& \textbf{\LO}\\
    \hline
    \multirow{2}*{Coverage range (km)}
    				   			& rural: $30$--$50$	& \multirow{2}*{$\approx 15$}
    				   												& rural: $10$--$15$\\
								& urban: $3$--$10$	& 				& urban: $3$--$5$\\
    \hline
    Frequency bands	(MHz)		& $868$ or $902$	& $2400$		& various, sub-GHz\\
	\hline
    ISM band					& \cmark			& \cmark		& \cmark\\
    \hline
    Bi-directional link			& \cmark			& \xmark		& \cmark\\
    \hline
    Data rate (Kbps)			& $0.1$				& $0.01$--$8$	& $0.3$--$37.5$\\
    \hline
    Nodes per BS				& $\approx 10^6$	& $\approx 10^4$& $\approx 10^4$\\
    \hline
   \end{tabular}
\end{table*}

%%%%%%%%%%%%%%%%%%%%%%%%%%%%%%%%
\section{Some Experimental Results Using a \LO Network}\label{experimental}
%%%%%%%%%%%%%%%%%%%%%%%%%%%%%%%%
In this section, we corroborate the argumentations of the previous sections by reporting some observations based on some initial deployments of \LO networks. 

\subsection{A \LO Deployment Test}
A \LO private network has been installed by Patavina Technologies s.r.l. in a large and tall building (19 floors) in Northern Italy, for a proof of concept of the capabilities of the \LO network. 
The objective is to monitor and control the temperature and the humidity of the different rooms, with the aim of reducing the costs related to heating, ventilation, and air conditioning. 
To this end, different wireless and wired communication technologies (including powerline communication) have been tried, but these solutions have been mostly unsatisfactory, requiring the installation of repeaters and gateways in basically every floor to guarantee mesh connectivity and access to the IP backbone.
Instead, the \LO technology has made it possible to provide the service by installing a single gateway on the ninth floor and placing 32 nodes all over the building, at least one per floor.
The installation included the integration of the NetServer with a monitoring application and with the databases already in use. 
At the time of writing, the installation has been flawlessly running for six months and is being considered as the preferred technology for the actual implementation of the energy saving program in many other buildings. 

We want to remark that the \LO network connectivity has been put under strain placing the nodes in elevators and in other places known to be challenging for radio connectivity.
All the stress tests have been successfully passed.
The envisioned next step is to install a gateway on an elevated site to serve multiple buildings in the neighborhood. %, since 5-6 \LVA{check\dots} buildings nearby are owned by the same administrator. 

This proof of concept is particularly relevant as it provides, on the one side, interesting insights on how pertinent and practical the LPWAN paradigm is for the Smart City scenario and, on the other side, some intuition from the economical point of view. Indeed, though extremely limited in its extent, the positive experience gained in the proof-of-concept installation of the \LO system in a building bodes well to the extension of the service to other public and private buildings, realizing at the same time an infrastructure for other Smart City services. According to Analysis Mason 2014 data, indeed, the number of LPWAN smart buildings connections is projected to be $0.8$ billions by 2023; and according to the McKinsey Global Institute analysis, the potential economic impact of IoT 2025 for Home and Cities is between \$1,1 and \$2,0 trillion. Thus, LPWAN solutions appear to have both the technical and the commercial capability to become the game changer in the Smart City scenario. 

\subsection{\LO Coverage Analysis}
One of the most debated aspects of LPWAN is the actual coverage range. This is crucial for a correct estimation of the costs for city-wide coverage, which may clearly have an important impact on the Capital Expenditure of the service providers. 

To gain insight in this respect, we carried out a coverage experimental test of \LO networks in the city of Padova, Italy.
The aim was to assess the ``worst case'' coverage of the technology, to have a conservative estimate of the number of gateways required to cover the whole city.
To this end, we placed a gateway with no antenna gain at the the top of a two storey building, without antenna elevation, in an area where high buildings are present. 

\begin{figure}[t!]
\begin{center}
\includegraphics[width=0.9\columnwidth]{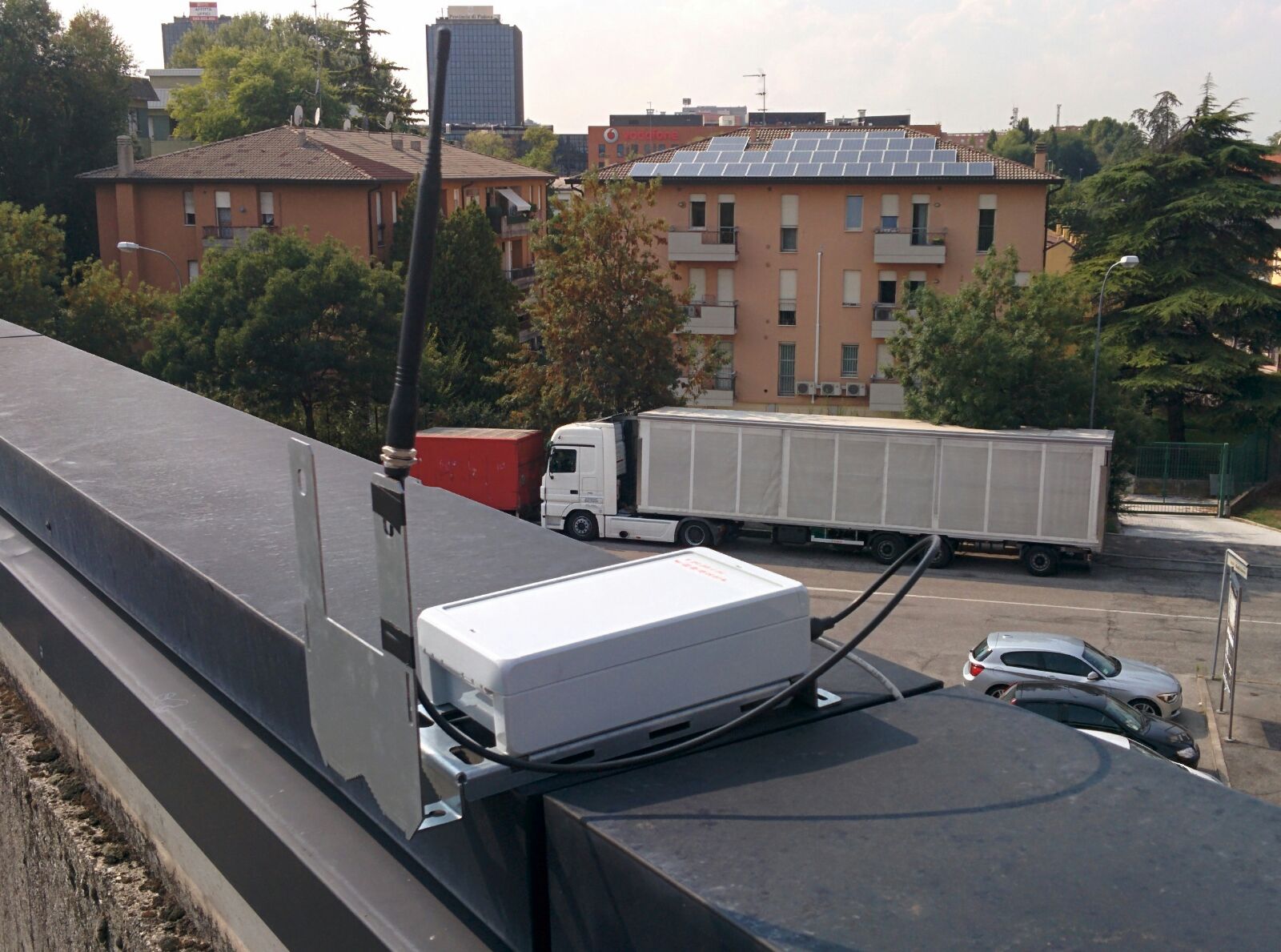}
\caption{\LO gateway installation.}
\label{accomodation}
\end{center}
\end{figure}

\begin{figure}[t!]
\begin{center}
\includegraphics[width=0.9\columnwidth]{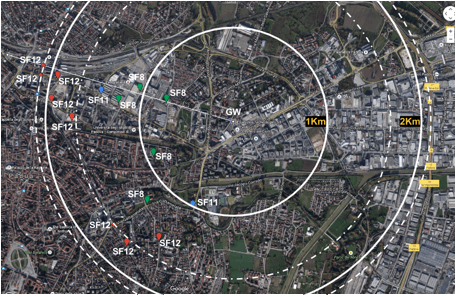}
\caption{\LO system single cell coverage in Padova, Italy. Worst case test.}
\label{cella}
\end{center}
\end{figure}

Fig.~\ref{accomodation} shows the experimental setup, while Fig.~\ref{cella} shows the results of the test. 
It can be seen that, in such harsh propagation conditions, the \LO technology allows to cover a cell of about 2~km of radius. 
However, the connection at the cell edge is guaranteed only when using the lowest bit rate (i.e., the longest spreading sequence which provides maximum robustness), with low margin for possible interference or to link budget changes. 
For this reason, we assumed a nominal coverage range of 1.2~km, a value that ensures a reasonable margin to interference and link budget variations due, e.g., to fading phenomena.

\begin{figure}[t!]
\begin{center}
\includegraphics[width=0.9\columnwidth]{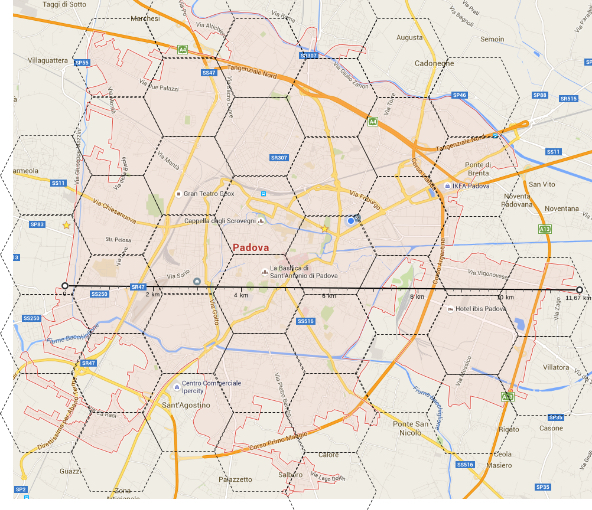}
\caption{\LO system cell coverage for Padova, Italy. Worst case test.}
\label{padova}
\end{center}
\end{figure}

Using this parameter, we attempted a rough coverage planning for the city of Padova, which extends over an area of about 100 square kilometers.
The resulting plan is shown in Fig.~\ref{padova}, from which we observe that, with the considered conservative coverage range estimate, the coverage of the entire municipality can be reached with a total of 30 gateways, which is less than half the number of sites deployed by one of the major cellular operators in Italy to provide mobile cellular access over the same area. 

Finally, we observe that Padova municipality accounts for about 200000 inhabitants. Considering 30 gateways to cover the city, we get about 7000 inhabitants per gateway.
The current \LO gateway technology claims the capability of serving 15000 nodes per gateway, which accounts for about 2 {\em things} per person.
Considering that the next generation of gateways is expected to triple the capacity (by using multiple directional antennas), in the long term we can expect that a basic coverage of the city may grant up to 6--7 {\em things} per person, on average, which seems to be more than adequate for most Smart City applications.
Any further increase in the traffic demand can be addressed by installing additional gateways, a solution similar to densification in cellular networks.

%\LVA{(i) una piccola sezione sull'installazione della RER (se ce la lasciano fare, ma non c'è nulla di segreto), in cui si dice che il sistema è sviluppato su più piani etc. e sta funzionando;} 
%
%\LVA{(ii) i risultati delle prove fatte per Padova, con figure di copertura della citt e magari qualche grafico di SNR (non so se ce l'abbiamo);} 
%
%\LVA{(iii) un'estrapolazione come l'esercizio fatto per la copertura di Padova in cui facciamo vedere i possibili posizionamenti dei gateway e i cerchi di copertura;} 
%
%\LVA{(iv) un conto approssimato della capacit di gestione di traffico, mostrando che un gateway che copre diversi km2 e' in grado di gestire il traffico generato in quell'area, finch la densit dei dispositivi e l'intensit di generazione del traffico rimangono entro certi limiti (che speriamo di far vedere che sono piuttosto ampi).}% Some experimental using a Lo-Ra network

%Conclusions
%%%%%%%%%%%%%%%%%%%%%%%%%%%%%%
\section{Conclusions}\label{conclusions}
%%%%%%%%%%%%%%%%%%%%%%%%%%%%%%

In this paper we have described the new emerging Low-Power Wide Area Networks (LPWANs) paradigm for Internet of Things connectivity.
This solution is based on long-range radio links, in the order of the tens of kilometers, and a star network topology, i.e., every node is directly connected to the base station.
Therefore, LPWANs are inherently different from usual IoT architectures, which are, instead, typically characterized by short-range links and a mesh topology.
The most prominent LPWAN technologies, i.e., \SF, \IN, and \LO, have been introduced and compared to the current short-range communication standards.
The experimental trials, which have been performed employing \LO technology, have shown that the LPWAN paradigm should be intended to complement current IoT standards as an enabler of Smart City applications, which can greatly benefit from long-range links.

\section*{Acknowledgment}
%%%%%%%%%%%%%%%%%%%%%%%%%%%%%%%%%%%%%%%%%%%%%%%%%
The authors would like to thank Ivano Calabrese and Nicola Bressan from Patavina Technologies, for their
contributions in the implementation of the \LO system and especially for running the field trials.

\ifCLASSOPTIONcaptionsoff
  \newpage
\fi

\begin{IEEEbiographynophoto}{Marco Centenaro}
(S'14) received the Bachelor's degree in Information Engineering in 2012 and the Master's degree in Telecommunication Engineering in 2014, both from the University of Padova, Italy.
Since November 2014 he is a Ph.D student at the Department of Information Engineering of the University of Padova, Italy.
His research interests include the next generation of cellular networks (5G) and in particular the Machine-to-Machine Communication. 
\end{IEEEbiographynophoto}

\begin{IEEEbiographynophoto}{Lorenzo Vangelista}
(S'93-M'97-SM'02) received the Laurea and Ph.D. degrees in electrical and telecommunication engineering from the University of Padova, Padova, Italy, in 1992 and 1995, respectively.
He subsequently joined the Transmission and Optical Technology Department, CSELT, Torino, Italy.
From December 1996 to January 2002, he was with Telit Mobile Terminals, Trieste, Italy, and then, until May 2003, he was with Microcell A/S, Copenaghen, Denmark. 
In July 2006, he joined the Worldwide Organization of Infineon Technologies as Program Manager. Since October 2006, he has been an Associate Professor of Telecommunication with the Department of Information Engineering, Padova University.
His research interests include signal theory, multicarrier modulation techniques, cellular networks and wireless sensors and actuators networks.\end{IEEEbiographynophoto}

\begin{IEEEbiographynophoto}{Andrea Zanella}
(S'98-M'01-SM'13) received the Laurea degree in computer engineering and Ph.D. degree in electronic and telecommunications engineering from the University of Padova, Padova, Italy, in 1998 and 2000, respectively.
He was a Visiting Scholar with the Department of Computer Science, University of California, Los Angeles (UCLA), Los Angeles, CA, USA, in 2000.
He is an Assistant Professor with the Department of Information Engineering (DEI), University of Padova.
He is one of the coordinators of the SIGnals and NETworking (SIGNET) research lab. His long-established research activities are in the fields of protocol design, optimization, and performance evaluation of wired and wireless networks.
\end{IEEEbiographynophoto}

\begin{IEEEbiographynophoto}{Michele Zorzi}
(S'89-M'95-SM'98-F'07) received his Laurea and PhD degrees in electrical engineering from the University of Padova in 1990 and 1994, respectively.
During academic year 1992-1993 he was on leave at UCSD, working on multiple access
in mobile radio networks.
In 1993 he joined the faculty of the Dipartimento di Elettronica e
Informazione, Politecnico di Milano, Italy.
After spending three years with the Center for Wireless Communications at UCSD, in 1998 he joined the School of Engineering of the University of Ferrara, Italy, where he became a professor in 2000.
Since November 2003 he has been on the faculty of the Information Engineering Department at the University of Padova.
His present research interests include performance evaluation in mobile communications systems, random access in wireless networks, ad hoc and sensor networks, Internet-of-Things, energy constrained communications protocols, cognitive networks, and underwater communications and networking.
He was Editor-In-Chief of IEEE Wireless Communications from 2003 to 2005 and Editor-In-Chief of the IEEE Transactions on Communications from 2008 to 2011, and is the founding Editor-In-Chief of the IEEE Transactions on Cognitive Communications and Networking.
He has also been an Editor for several journals and a member of the Organizing or the Technical Program Committee for many international conferences, as well as guest editor for special issues in IEEE Personal Communications, IEEE Wireless Communications, IEEE Network and IEEE Journal on Selected Areas in Communications.
He served as a Member-at-Large of the Board of Governors of the IEEE Communications Society from 2009 to 2011, and is currently its Director of Education and Training.
\end{IEEEbiographynophoto}


%\vfill



\begin{thebibliography}{10}
\providecommand{\url}[1]{#1}
\csname url@samestyle\endcsname
\providecommand{\newblock}{\relax}
\providecommand{\bibinfo}[2]{#2}
\providecommand{\BIBentrySTDinterwordspacing}{\spaceskip=0pt\relax}
\providecommand{\BIBentryALTinterwordstretchfactor}{4}
\providecommand{\BIBentryALTinterwordspacing}{\spaceskip=\fontdimen2\font plus
\BIBentryALTinterwordstretchfactor\fontdimen3\font minus
  \fontdimen4\font\relax}
\providecommand{\BIBforeignlanguage}[2]{{%
\expandafter\ifx\csname l@#1\endcsname\relax
\typeout{** WARNING: IEEEtran.bst: No hyphenation pattern has been}%
\typeout{** loaded for the language `#1'. Using the pattern for}%
\typeout{** the default language instead.}%
\else
\language=\csname l@#1\endcsname
\fi
#2}}
\providecommand{\BIBdecl}{\relax}
\BIBdecl

\bibitem{M2M1}
C.~Anton-Haro and M.~Dohler, \emph{{Machine-to-Machine (M2M) Communications:
  Architecture, Performance and Applications}}, 1st~ed.\hskip 1em plus 0.5em
  minus 0.4em\relax Woodhead Publishing Ltd., Jan. 2015.

\bibitem{IoT1}
\BIBentryALTinterwordspacing
J.~Gubbi, R.~Buyya, S.~Marusic, and M.~Palaniswami, ``{Internet of Things
  (IoT): A Vision, Architectural Elements, and Future Directions},''
  \emph{Future Generation Computer Systems}, vol.~29, no.~7, pp. 1645--1660,
  Sept. 2013. [Online]. Available:
  \url{http://dx.doi.org/10.1016/j.future.2013.01.010}
\BIBentrySTDinterwordspacing

\bibitem{IoT2}
\BIBentryALTinterwordspacing
D.~Miorandi, S.~Sicari, F.~De~Pellegrini, and I.~Chlamtac, ``{Internet of
  Things},'' \emph{Ad Hoc Networks}, vol.~10, no.~7, pp. 1497--1516, Sept.
  2012. [Online]. Available:
  \url{http://dx.doi.org/10.1016/j.adhoc.2012.02.016}
\BIBentrySTDinterwordspacing

\bibitem{IoT3}
A.~Zanella, N.~Bui, A.~Castellani, L.~Vangelista, and M.~Zorzi, ``{Internet of
  Things for Smart Cities},'' \emph{IEEE Internet of Things Journal}, vol.~1,
  no.~1, pp. 22--32, Febr. 2014.

\bibitem{M2M4}
\BIBentryALTinterwordspacing
A.~Biral, M.~Centenaro, A.~Zanella, L.~Vangelista, and M.~Zorzi, ``{The
  challenges of M2M massive access in wireless cellular networks},''
  \emph{Digital Communications and Networks}, vol.~1, no.~1, pp. 1 -- 19, Feb.
  2015. [Online]. Available:
  \url{http://www.sciencedirect.com/science/article/pii/S235286481500005X}
\BIBentrySTDinterwordspacing

\bibitem{M2M2}
S.-Y. Lien, K.-C. Chen, and Y.~Lin, ``{Toward Ubiquitous Massive Accesses in
  3GPP Machine-to-Machine ommunications},'' \emph{IEEE Communications
  Magazine}, vol.~49, no.~4, pp. 66--74, Apr. 2011.

\bibitem{M2M3}
3GPP, ``{Standardization of Machine-Type Communications},'' Tech. Rep. V0.2.2,
  Dec. 2013.

\bibitem{CIoT}
{Vodafone Group Plc.}, ``{New Study Item on Cellular System Support for Ultra
  Low Complexity and Low Throughput Internet of Things},'' 3GPP TSG GERAN\#62,
  Tech. Rep. GP-140421, May 2014.

\bibitem{etsi300}
ETSI, ``{Electromagnetic compatibility and Radio spectrum Matters (ERM)},''
  Tech. Rep. 300 220-1 V2. 4.1, Jan. 2012.

\bibitem{8021514k}
``{IEEE Standard for Local and metropolitan area networks -- Part 15.4:
  Low-Rate Wireless Personal Area Networks (LR-WPANs) -- Amendment 5: Physical
  Layer Specifications for Low Energy, Critical Infrastructure Monitoring
  Networks},'' \emph{IEEE Standard 802.15.4k-2013}, pp. 1--149, Aug. 2013.

\bibitem{onramp-patent}
\BIBentryALTinterwordspacing
{On-Ramp Wireless Inc.}, ``Light monitoring system using a random phase
  multiple access system,'' July 2013, {US Patent 8,477,830}. [Online].
  Available: \url{https://www.google.com/patents/US8406275}
\BIBentrySTDinterwordspacing

\bibitem{Sforza2013}
\BIBentryALTinterwordspacing
F.~Sforza, ``Communications system,'' Mar. 2013, {US Patent 8,406,275}.
  [Online]. Available: \url{https://www.google.com/patents/US8406275}
\BIBentrySTDinterwordspacing

\bibitem{Berni1973}
A.~J. Berni and W.~Gregg, ``{On the Utility of Chirp Modulation for Digital
  Signaling},'' \emph{IEEE Transactions on Communications}, vol.~21, no.~6, pp.
  748--751, June 1973.

\bibitem{etsi300220}
ETSI, ``{Electromagnetic compatibility and Radio spectrum Matters (ERM); Short
  Range Devices (SRD); Radio equipment to be used in the 25 MHz to 1 000 MHz
  frequency range with power levels ranging up to 500 mW; Part 1: Technical
  characteristics and test methods},'' Tech. Rep. EN 300 220-1 V2.4.1, Jan.
  2012.

\bibitem{lorawan}
{LoRa\textsuperscript{\texttrademark} Alliance},
  ``{LoRa\textsuperscript{\texttrademark} Specification V1.0},'' Tech. Rep.,
  May 2015.

\end{thebibliography}
\end{document}